\documentclass[twocolumn,showpacs,preprintnumbers,amsmath,amssymb,widetext,super
scriptaddress]{revtex4}

\usepackage{graphicx}

\begin{document}

\title{Mesoscopic supercurrent transistor controlled by nonequilibrium cooling}
\author{F. Giazotto}
\email{giazotto@sns.it} \affiliation{NEST-INFM \& Scuola Normale
Superiore, I-56126 Pisa, Italy}
\author{T. T. Heikkil\"{a}}
\affiliation{Low Temperature Laboratory, Helsinki University of
Technology, P.O. Box 2200, FIN-02015 HUT, Finland}
\author{F. Taddei}
\affiliation{NEST-INFM \& Scuola Normale Superiore, I-56126 Pisa,
Italy}
\author{Rosario Fazio}
\affiliation{NEST-INFM \& Scuola Normale Superiore, I-56126 Pisa,
Italy}
\author{J. P. Pekola}
\affiliation{Low Temperature Laboratory, Helsinki University of
Technology, P.O. Box 2200, FIN-02015 HUT, Finland}
\author{F. Beltram}
\affiliation{NEST-INFM \& Scuola Normale Superiore, I-56126 Pisa,
Italy}

\begin{abstract}
The distinctive quasiparticle  distribution  existing under
nonequilibrium in a superconductor-insulator-normal
metal-insulator-superconductor (SINIS) mesoscopic line is proposed
as a novel tool to  control  the supercurrent intensity in a long
Josephson weak link. We present a description of this system in
the framework of the diffusive-limit quasiclassical Green-function
theory and take into account the effects of inelastic scattering
with arbitrary strength. Supercurrent enhancement and suppression,
including a marked transition to a $\pi$-junction are striking
features leading to a fully tunable structure. The role of the
degree of nonequilibrium, temperature, and materials choice as
well as features like noise, switching time, and current and power
gain are also addressed.
\end{abstract}
\pacs{74.50.+r, 73.23.-b, 74.40.+k}

\maketitle

\section{INTRODUCTION}
Nonequilibrium effects in mesoscopic  superconducting circuits
have been receiving a rekindled attention during the  last few
years \cite{articles}. The art of controlling Josephson coupling
in superconductor-normal metal-superconductor (SNS) weak links is
at present  in the spotlight: a recent breakthrough in mesoscopic
superconductivity is indeed represented by the  SNS transistor,
where supercurrent suppression  as well as its sign reversal
($\pi$-transition) were  demonstrated
\cite{baselmans,baselmansprb,huang,tak,baselmansphase}. This was
achieved by driving the quasiparticle distribution in the weak
link far from equilibrium \cite{volkov,wilhelm,yip} through
external voltage terminals, \emph{viz.} normal reservoirs. Such a
behavior relies on
 the two-step shape of the quasiparticle nonequilibrium
 distribution, typical of diffusive mesoscopic wires and
experimentally observed by Pothier and coworkers \cite{pothier}.

The purpose of this paper is to demonstrate that  it is possible
 to tailor the quasiparticle distribution through
\emph{superconductivity-induced} nonequilibrium in order to
implement a unique class of superconducting transistors
\cite{SINISTrans}. This can be achieved
 when mesoscopic control lines are connected to
superconducting reservoirs through tunnel barriers (I), realizing
a SINIS channel. The peculiar  quasiparticle
 distribution  in the N region, originating  from
 biasing the S terminals, allows one to access several regimes,  from supercurrent
enhancement with respect to equilibrium   to a large amplitude of
the $\pi$-transition passing through a steep supercurrent
suppression. These features are accompanied by a large current
gain (up to some $10^{5}$ in the  region of larger input
impedance) and reduced dissipation. The ultimate operating
frequencies available  open the way to the exploitation of this
scheme for the implementation of  ultrafast cryogenic current
and/or power amplifiers.

The paper is organized as follows. In Sec. II we introduce
diffusive-limit quasiclassical Green-function theory that is
employed to describe the supercurrent in the proposed structure.
In particular, we show how to exploit superconductivity-induced
nonequilibrium in order to control the Josephson current. In Sec.
III we address the role of inelastic scattering in the structure
by solving the kinetic equation for the SINIS control line, and we
describe how the strength  of such scattering affects the
observable supercurrent. The critical current  dependence on the
length of the weak link is analyzed in Sec. IV where the relevance
of \emph{long} junctions is pointed out. In Sec. V we consider the
effect of the additional control terminals  on the observable
supercurrent while Sec. VI is devoted to the analysis of the
supercurrent harmonics. In Sec. VII we address the role of
materials combination in determining the performance of the
structure. In particular, issues like power dissipation and noise
power are discussed. The achievable current and power gain as well
as the maximum attainable transistor operating frequency are
discussed in Sec. VIII. Finally, Sec. IX summarizes the main
results.

\begin{figure}[ht!]
\begin{center}
\includegraphics[width=8.0cm,clip]{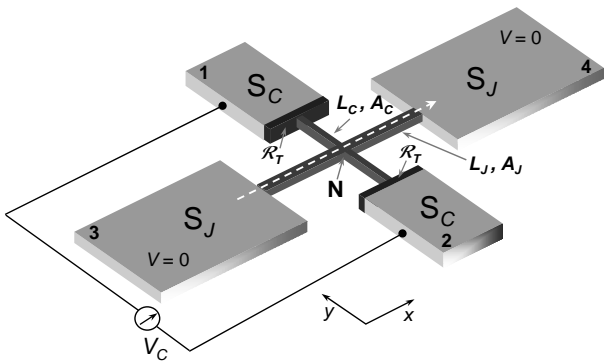}
\end{center}
\caption{Scheme of the Josephson transistor. The supercurrent
$I_J$ (along the white dashed line) is tuned by applying a bias
$V_C$ across the SINIS symmetric  line connected to the center of
the weak link. The superconducting gaps $\Delta _J$ and $\Delta
_C$ are, in general, different, and all normal wires are assumed
quasi-one-dimensional, i.e., their width is much smaller than
their length.} \label{transistor}
\end{figure}

\section{TRANSISTOR OUTPUT CHARACTERISTIC}
The investigated mesoscopic structure  (see Fig.~\ref{transistor})
consists of a long diffusive weak link of length $L_J$ much larger
than the superconducting coherence length ($\xi_0$) oriented along
the $x$ direction. This defines the SNS junction of cross-section
$A_J$. The superconducting terminals belonging to the SNS
junction, labelled S$_J$ (3 and 4), are kept at zero potential in
the analysis of the supercurrent (transistor operation may take
place in the dissipative regime at a finite voltage). The SINIS
control line is oriented along the $y$ direction and consists  of
a normal wire, of length $L_C$ and cross-section $A_C$, connected
through identical tunnel junctions of resistance $\mathcal{R}_T$
to two superconducting reservoirs S$_C$ (1 and 2), biased at
opposite voltages $\pm V_C /2$. The superconducting gaps of S$_J$
and S$_C$ ($\Delta _J$ and $\Delta _C$) are in general different
in order to optimize the structure characteristics.

The supercurrent $I_J$ flowing across the SNS junction is  given
by \cite{wilhelm,yip}
\begin{equation}
I_J(V_C)=\frac{\sigma A_J}{eL_J}\int_0^\infty dE\,
[f(-E;V_C)-f(E;V_C)]\textrm{Im}[j_E], \label{supercurrent}
\end{equation}
and depends on the quasiparticle distribution function $f(E)$ in
N. In Eq.~(\ref{supercurrent}), $\sigma$ is the normal-state
conductivity which determines the normal-state resistance of the
junction according to $R_N =L_J /\sigma A_J$. As shown below, the
distribution function $f$ is determined by the SINIS control line
and reduces to the equilibrium Fermi distribution when
 $V_C =0$.

The energy-dependent spectral supercurrent, \cite{belzig,tero}
${\rm Im}[j_E]$, can be calculated by solving the Usadel equations
\cite{usadel}. Following the parametrization of the Green
functions given in Ref.~\onlinecite{belzig}, these equations in
the N region can be written as
\begin{eqnarray}
j_E =-\sinh ^2(\theta)\partial_x\chi,\,\,\,\,\,\,\,\,\,\,\,
\partial_xj_E=0,\\
\hbar D\partial_x ^2\theta+2iE\sinh\theta-\frac{\hbar
D}{2}(\partial_x\chi)^2\sinh(2\theta)=0,
 \label{Usadel}
\end{eqnarray}
where $D=\frac{1}{3}v_F\ell_m$ is the diffusion coefficient, $v_F$
is the  Fermi velocity, $\ell _m$ is the mean free path and $E$ is
the energy relative to the chemical potential in S$_J$.
$\theta(x,E)$ and $\chi(x,E)$ are in general complex functions.
\begin{figure}[ht!]
\begin{center}
\includegraphics[width=8.0cm,clip]{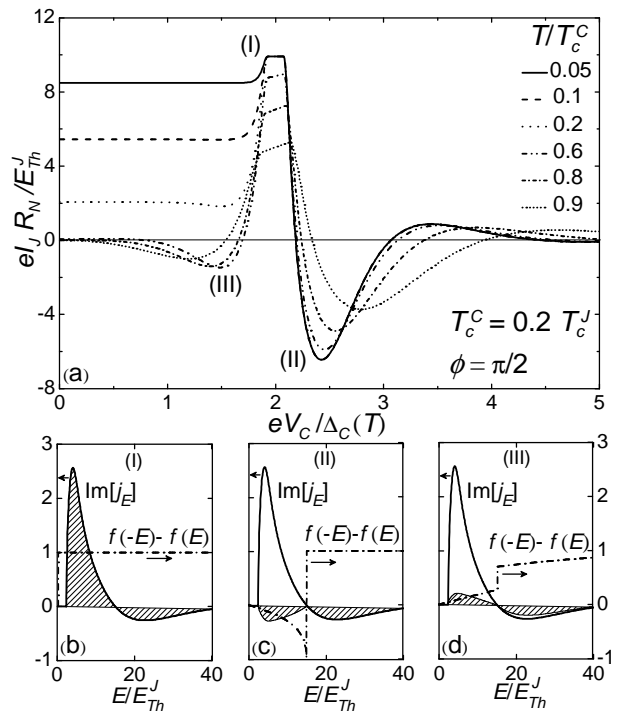}
\end{center}
\caption{(a) Supercurrent  vs control voltage $V_C$ at different
temperatures ($T$) for $\phi =\pi /2$ and
$T_{c}^{C}=0.2\,T_{c}^{J}$ (see text). Bias regions labelled (I),
(II) and (III) indicate supercurrent enhancement due to
quasiparticle \emph{cooling}, high-voltage $\pi$-state and
low-voltage $\pi$-state in the high-temperature regime,
respectively. These are qualitatively explained in (b), (c) and
(d) where hatched areas represent the contribution to supercurrent
arising in such bias ranges (see text). We assume throughout that
supercurrent at $\phi=\pi/2$ yields values close enough to the
critical current, an assumption that breaks down near the
$\pi$-junction transition \cite{tero,baselmansphase} (see
Sec.~\ref{sec:harmonics}).} \label{supercurrentbehavior}
\end{figure}
For perfectly transmissive contacts, the boundary conditions at
the S$_J$N interfaces reduce to $\theta=\,$arctanh$(\Delta_J/E)$
and $\chi=\pm\phi/2$ in the reservoirs $S_J$, where $\phi$ is the
phase difference between the superconductors. The presence of the
control wires on the spectral supercurrent can be taken into
account explicitly including two additional probes and imposing
boundary conditions at the S$_C$N interfaces as in Ref.
[\onlinecite{tero}]. In the present case such boundary conditions
take the form $\partial_x \theta=0$ to describe large tunnel
barriers of resistance ${\mathcal R}_T$. The large ${\cal R}_T$
allows us to neglect the tiny oscillating supercurrents flowing
between the superconductors $S_C$. As discussed in Sec. V, the
supercurrent $I_J$ depends trivially on the control wire length
$L_C$, namely it monotonically decreases, rapidly reaching an
asymptotic value of, roughly, one half the zero-length value. This
in turn allows us to make the following simplifications: as far as
the spectral current is concerned, we shall assume
$L_C/L_J\rightarrow 0$, aware of the fact that the effect of a
finite $L_C/L_J$ gives rise to a mere reduction of $I_J$. A
similar discussion can be made on the dependence of the
equilibrium supercurrent on $A_C$. As shown in Ref.
[\onlinecite{tero}], $I_J$ is a monotonically decreasing function
of $A_C/A_J$ and, as far as spectral supercurrent is concerned, we
assume $A_C/A_J \rightarrow 0$ throughout the paper (in fact, it
is enough to assume either $L_C \ll L_J$ or $A_C \ll A_J$).

We must now determine the actual quasiparticle distribution $f$ in
the N region of the SINIS structure. This is controlled by voltage
($V_C$) and temperature, and  by the amount of inelastic
scattering in the control line. In the case of a short control
wire with no inelastic interactions, the quasiparticle
distribution, according to Ref. [\onlinecite{Heslinga}], is given
by
\begin{equation}
f(E,V_C)=\frac{\mathcal{N}_1 \mathcal{F}_1 +\mathcal{N}_2
\mathcal{F}_2}{\mathcal{N}_1 +\mathcal{N}_2},
 \label{noneq}
\end{equation}
where $\mathcal{N}_{1,2}=\mathcal{N}_{S_C}(E\pm eV_C/2)$ and
$\mathcal{F}_{1,2}=\mathcal{F}^0(E\pm eV_C/2 )$. The former are
the BCS density of states in the reservoirs S$_C$ (labelled 1 and
2 in Fig.~\ref{transistor}). $\mathcal{F}^0(E)$ is the Fermi
function at lattice temperature $T$ \cite{lowenote}. Note that
expression (\ref{noneq}) can also be found from the quasiclassical
theory in the tunnelling limit \cite{brinkman03}. In this case
Eqs. (\ref{supercurrent}) and (\ref{noneq}) yield the
dimensionless transistor output characteristics shown in Fig.
\ref{supercurrentbehavior}(a), where we plot the supercurrent
$I_J$ versus control bias $V_C$ at different temperatures. We
assumed $\phi=\pi/2$, $T_c ^C /T_c ^J =0.2$, where $T_c ^{C(J)}$
are the critical temperatures of the superconductors S$_{C(J)}$,
and $L_J$ such that $\Delta_J /E_{Th}^J =300$. We choose the limit
of a long junction (i.e., $\Delta_J \gg E_{Th}^J$, where $E_{Th}^J
=\hbar D/L_J ^2$ is the Thouless energy of the SNS junction),
since in this limit the supercurrent spectrum varies strongly with
energy.

For all temperatures $T < T_c^C$, with increasing $V_C$, curves
display a large supercurrent enhancement with respect to
equilibrium at bias $V_C = 2\,\Delta_C (T)/e=V_C ^\ast (T)$
(region I in the figure). Further increase of the bias leads to a
$\pi$-transition (region II) and finally to a decay for larger
voltages \cite{baselmansthesis}. This behavior is illustrated by
Figs.~2(b,c,d) where the spectral supercurrent (solid line) is
plotted together with $f(-E)-f(E)$ (dash-dotted line) for values
of $V_C$ and $T$ corresponding to regions I, II and III,
respectively. Hatched areas represent the integral of their
product, i.e., the supercurrent $I_J$ of Eq. (\ref{supercurrent}).
In particular, region I corresponds to the \emph{cooling} regime
where hot quasiparticles are extracted from the normal metal
\cite{Heslinga,leivo}. The origin of the $\pi$-transition in
region II is illustrated by Fig.~\ref{supercurrentbehavior}(c),
where the negative contribution to the integral is shown.  We
remark that the intensity of the supercurrent inversion is very
significant. It reaches about 60$\%$ of the maximum value of $I_J$
at $V_C \simeq V_C ^\ast (T)$ in the whole temperature range,
nearly doubling the $\pi$-state value of the supercurrent as
compared to the case of an all-normal control channel
\cite{wilhelm,yip}.
 In the high-temperature regime ($T/T_c^C\geq 0.6$), when the equilibrium
critical current is vanishing, the supercurrent first undergoes a
low-bias $\pi$-transition (region III in the figure), then enters
regions I and II. This recovery of the supercurrent from
vanishingly small values at equilibrium is again a consequence of
the peculiar shape of $f$ (see
Fig.~\ref{supercurrentbehavior}(d)). Notably, the supercurrent
enhancement around $V_C ^\ast(T)$ remains  pronounced even at the
highest temperatures, so that $I_J$ attains values largely
exceeding 50$\%$ of the junction maximum supercurrent.
 This demonstrates the
strong tunability  of the supercurrent through nonequilibrium
effects induced by the superconducting control lines. We remark
that this is a unique feature stemming from the
superconductivity-induced nonequilibrium population in the weak
link.

\section{ROLE OF INELASTIC SCATTERING} Since we do not want to
confine our analysis only to the limit of control wires much
shorter than the energy relaxation length, we need to evaluate the
impact of inelastic scattering present in the N region on the
supercurrent behavior \cite{intspec}. As a matter of fact, the
length $L_C$ of the SINIS control line can be additionally varied
to control the supercurrent by changing the effective strength of
inelastic scattering in the N region. In general, the steady state
distribution function at a given energy $E$ is determined by an
equation which accounts for the balancing of quasiparticle flow to
and from the reservoirs with a collision term $\mathcal{I}$,
describing energy relaxation processes. For $\mathcal{R}_T \gg R_C
=L_C /\sigma A_C$, the distribution function $f(E)$ in the N
region is essentially $y$-independent and we have
\begin{equation}
\begin{split}
\frac{1}{e^2 \mathcal{R}_T \Omega_C
\nu_F}[\mathcal{N}_1(\mathcal{F}_1-f(E))+
\mathcal{N}_2(\mathcal{F}_2-f(E))]+\\+\kappa \int d\omega
d\varepsilon \,\omega^{\alpha}
\mathcal{I}(\omega,\varepsilon,E)=0.
\end{split}
\label{kinetic}
\end{equation}
Here $\nu_F$ is the normal-metal density of states at the Fermi
energy, $\Omega_C$ is the volume of the N region and $\mathcal{I}$
is the net collision rate at energy $E$. At low temperatures, the
most relevant scattering mechanism is electron-electron scattering
\cite{anthore} and  we can neglect the effect of electron-phonon
scattering. Then, \cite{nag,pothier}
\begin{equation}
\mathcal{I}(\omega,\varepsilon,E)=\mathcal{I}^{in}(\omega,\varepsilon,E)-\mathcal{I}^{out}(\omega,\varepsilon,E),
\label{collision1}
\end{equation}
and
\begin{eqnarray}
\mathcal{I}^{in}(\omega,\varepsilon,E)=(1-f({\varepsilon}))(1-f(E))f({\varepsilon-\omega})
f(E+\omega),\\
\mathcal{I}^{out}(\omega,\varepsilon,E)=(1-f({\varepsilon-\omega}))(1-f({E+\omega}))f({\varepsilon})f(E).
\label{collision2}
\end{eqnarray}
Electron-electron interaction is either due to direct Coulomb
scattering \cite{AA,kamenev} or mediated by magnetic impurities
\cite{anthore}. Below, we concentrate on the former, but the
latter would yield  a similar qualitative behavior. From the
calculation of the screened Coulomb interaction in the diffusive
channel, it follows \cite{AA} that $\alpha =-3/2$ for a quasi-one
dimensional wire and $\kappa
=(\pi\sqrt{2D}\hbar^{3/2}\nu_{F}A_C)^{-1}$ \cite{kamenev,huard}.
We note that $\Delta_C$ is the most relevant energy scale  to
describe  the distribution function for different voltages $V_C$.
It is thus useful to replace $\omega\rightarrow \omega/\Delta_C$
and $\varepsilon\rightarrow \varepsilon/\Delta_C$, in order to
obtain a dimensionless equation. Multiplying Eq.~(\ref{kinetic})
by $e^2 \mathcal{R}_T \Omega_C \nu_F$ we obtain
\begin{equation}
\begin{split}
\mathcal{N}_1(\mathcal{F}_1 -f(E))-
\mathcal{N}_2(f(E)-\mathcal{F}_2)=\\=\mathcal{K}_{coll} \int
d\omega d\varepsilon \omega^{-3/2}
\mathcal{I}(\omega,\varepsilon,E),
\end{split}
\label{SINISdistribution}
\end{equation}
where
\begin{equation}
\mathcal{K}_{coll}=\frac{\mathcal{R}_T}{R_C}\frac{L_C^2\kappa}{D}\sqrt{\Delta_C}=\frac{1}{\sqrt{2}}\frac{\mathcal{R}_T}{R_K}\sqrt{\frac{\Delta_C}{E_{Th}^C}},
\end{equation}
$R_K =h/2e^2$ and $E_{Th}^C =\hbar D/L_C ^2$. In the absence of
electron-electron interaction ($\mathcal{K}_{coll}=0$), Eq.
(\ref{noneq}) is  recovered.
\begin{figure}[h!]
\begin{center}
\includegraphics[width=8.0cm,clip]{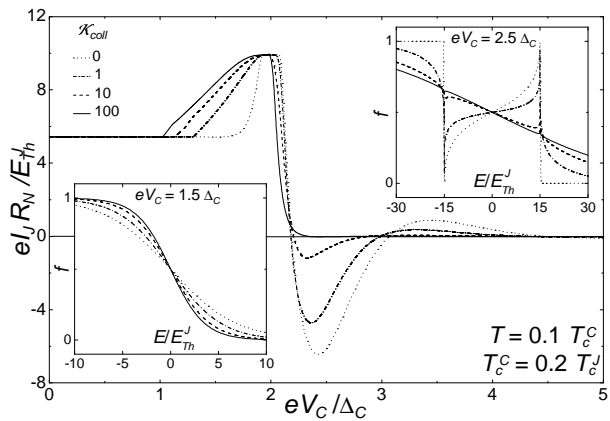}
\end{center}
\caption{Supercurrent  vs  $V_C$ for various $\mathcal{K}_{coll}$
with $T=0.1\,T_c ^C$ and $T_{c}^{C}=0.2\,T_{c}^{J}$. Insets show
the distribution function at
 $eV_C =1.5\Delta_C$ (left) and  $eV_C
=2.5\Delta_C$  (right) calculated for the same
$\mathcal{K}_{coll}$ values.} \label{SINISdistrfig}
\end{figure}

The influence of inelastic scattering on $I_J$ is shown in
Fig.~\ref{SINISdistrfig}, which displays the critical current of a
long junction  at $T=0.1\,T_c ^C$ for several values of
$\mathcal{K}_{coll}$. Here $I_J$ is obtained by numerically
solving Eq. (\ref{SINISdistribution}). The effect of
electron-electron interaction is to strongly suppress the
$\pi$-state and to widen the
 peak around $V_C ^\ast$. The $\pi$-transition
 vanishes for $\mathcal{K}_{coll}\simeq 100$, but
the $I_J$ enhancement due to quasiparticle cooling still persists
in the limit of even larger inelastic scattering \cite{giazotto}.

The disappearance of the $\pi$-state can be understood by looking
at the right inset of Fig.~\ref{SINISdistrfig} which clearly shows
how a large-bias $f$ (calculated at $eV_C =2.5\Delta_C$) gradually
relaxes from nonequilibrium towards a Fermi  function upon
increasing $\mathcal{K}_{coll}$. The left inset shows how a
low-bias $f$ (evaluated at $eV_C =1.5\Delta_C$) sharpens, thus
enhancing $I_J$, by increasing $\mathcal{K}_{coll}$. This effect
follows from the fact that inelastic interactions redistribute the
occupation of quasiparticle levels in the N region, thus
increasing the occupation
 at higher energy. As a consequence, higher-energy
excitations are more efficiently
 removed  by tunneling, even for biases well below and
not only around $V_C ^\ast$ (as in the case of
$\mathcal{K}_{coll}=0$). At the same time, supercurrent recovery
at high temperature is gradually weakened upon enhancing
$\mathcal{K}_{coll}$. Notably, these calculations show that a
rather large amount of inelastic scattering is necessary to weaken
and completely suppress the $\pi$-state. For example, using
Al/Al$_2$O$_3$/Cu as materials composing the SINIS line,
$\mathcal{K}_{coll}= 1$ corresponds to use a fairly long control
line with $L_C\simeq 4.6$ $\mu$m \cite{device}.

\begin{figure}[ht!]
\begin{center}
\includegraphics[width=8.0cm,clip]{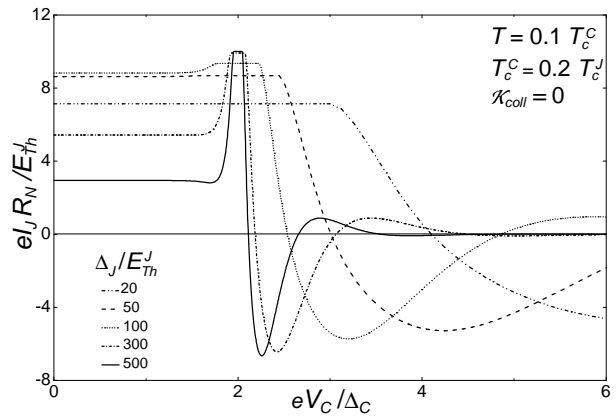}
\end{center}
\caption{Effect of different ratios $\Delta _J /E_{Th} ^J$ on the
observable supercurrent at $T=0.1\,T_c ^C$ and
$T_{c}^{C}=0.2\,T_{c}^{J}$. Note that the effect of the
nonequilibrium distribution on the observable supercurrent is more
pronounced in longer junctions.} \label{fig:length}
\end{figure}
\section{DEPENDENCE ON JUNCTION LENGTH} \label{sec:ldep} The choice of the
superconductors S$_J$ and S$_C$ affects the behavior of the device
through the energy scales given by the energy gaps $\Delta_{J/C}$
in these materials and their ratio with the geometric scales of
the system. The role of junction length ($L_J$) on the transistor
characteristics is displayed in Fig. \ref{fig:length}, where the
supercurrent is plotted for different ratios $\Delta _J /E_{Th}
^J$ at $T=0.1\,T_c ^C$ and for $T_{c}^{C}=0.2\,T_{c}^{J}$. Upon
shortening the junction the Josephson current response to the
control voltage $V_C$ becomes wider (note that also the ratio
between the characteristic energies $\Delta_C$ and $E_{Th}^J$
changes --- our model holds only for $\Delta_J \gtrsim \Delta_C$).
At the same time, features like supercurrent enhancement and
$\pi$-transition appear less pronounced. This suggests that large
$\Delta _J /E_{Th} ^J$ ratios are favorable for an efficient
transistor effect, so that the $I_J (V_C)$ characteristic is
forced into a narrower bias window.

\section{EFFECT OF EXTRA TERMINALS} 
\begin{figure}[h!]
\begin{center}
\includegraphics[width=8.0cm,clip]{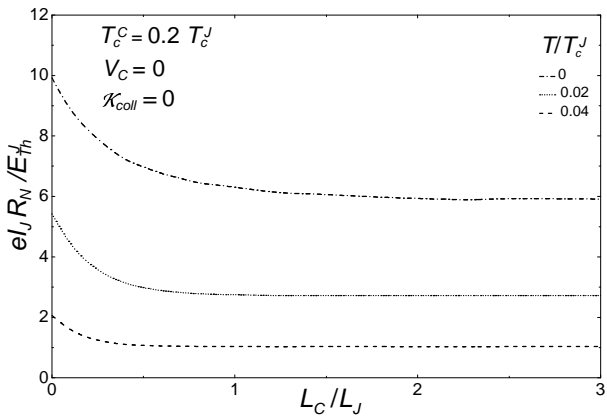}
\end{center}
\caption{Equilibrium supercurrent vs $L_C /L_J$  at different
temperatures. Note that $I_J$ reaches its asymptotic value at $L_C
\simeq L_J$, almost independently of the temperature.}
\label{fig:terminals}
\end{figure}

As mentioned earlier,
the presence of the lateral arms influences the superconducting
correlations induced in the weak link  by reducing \cite{tero} the
spectral supercurrent. Such an effect is shown in
Fig.~\ref{fig:terminals} where the equilibrium supercurrent (i.e.,
at $V_C =0$) is plotted as a function of the ratio $L_C/L_J$. The
magnitude of the supercurrent is decreased upon increasing $L_C$,
reaching its asymptotic value at $L_C \simeq L_J$, almost
independently of temperature. Note that, for each temperature,
such a value is roughly one half that corresponding to the limit
$L_C\rightarrow 0$. This conclusion contrasts with the case of a
normal control line without tunnel barriers \cite{yip,tero} and
emphasizes the relevance of the condition $L_C \ll L_J$ towards
the maximization of the supercurrent in the system.

\section{PHASE DEPENDENCE}\label{sec:harmonics}
The generic sinusoidal supercurrent-phase dependence is strictly
valid only in the case when the Josephson junction is a tunnel
junction. Transmission channels with high transmission
probability, present in diffusive wires, tend to deform this
dependence by bringing in higher harmonics. Thus, the
supercurrent-phase relation is in general of the form \cite{tero}
\begin{equation}
I_S(\phi)=\sum_{n=1}^\infty I_{J,n}\sin(n\phi),
\end{equation}
where $I_{J,n}$ are the coefficients of the Fourier sine series of
$I_S(\phi)$. Loosely speaking, the higher harmonics arise from
multiple passages of Cooper pairs through the junction in a single
coherent process. In most cases for a diffusive junction, the
$n=1$ harmonic is dominant, and the role of the others is to
slightly deform the current-phase relation, without bringing about
additional nodes. However, across the $\pi$-junction transition,
the first harmonic can be completely suppressed. In this case, the
current-phase relation is determined by the second harmonic, and
around this point, the supercurrent has a half periodicity with
respect to the phase (flux in a loop). Such a $\sin(2\phi)$ phase
dependence was measured in Ref. [\onlinecite{baselmansphase}] in
the case of normal-metallic control probes.
\begin{figure}[h!]
\begin{center}
\includegraphics[width=8.0cm,clip]{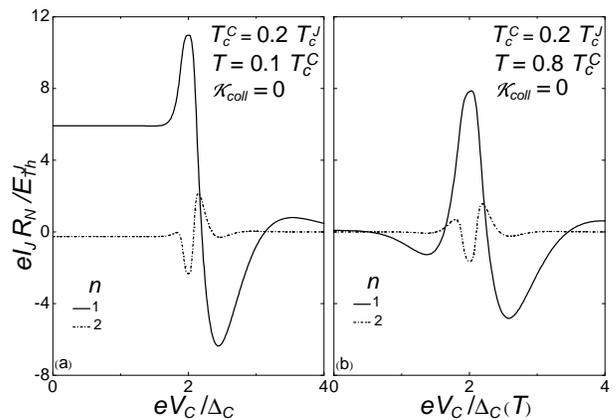}
\end{center}
\caption{(a) Voltage dependence of the amplitude of the first two
harmonics of the observable supercurrent calculated at $T=0.1\,T_c
^C$ and $T_{c}^{C}=0.2\,T_{c}^{J}$. (b) The same as in (a)
calculated in the high-temperature regime ($T=0.8\,T_c ^C$).}
\label{fig:harmonics}
\end{figure}

Figure \ref{fig:harmonics} shows the voltage dependence of the two
lowest harmonics at two temperatures. One can find that for most
voltages, the first harmonic is dominant, but in the region(s) of
the transition between the conventional and the $\pi$-states, the
second becomes dominant. Moreover, one finds that, analogously to
the case with normal control probes, \cite{baselmansphase,tero}
the sign of the second harmonic in these regions is positive.
Hence, in those values of the voltage $V_C$, the free energy of
the junction has local minima both at $\phi=0$ and $\phi=\pi$
\cite{baselmansphase}. Another implication of the finite value of
$I_{J,2}$ around those voltages is that the critical current of
the junction does not completely vanish, but it is merely no
longer obtained for $\phi \approx \pi/2$.

\begin{figure}[hb!]
\begin{center}
\includegraphics[width=8.0cm,clip]{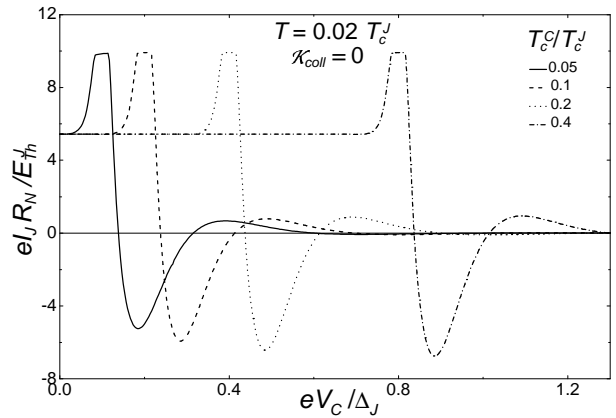}
\end{center}
\caption{Supercurrent vs $V_C$ calculated at $T=0.02\,T_c^J$ for
different $T_c^C /T_c^J$ ratios. The main effect of lowering
$T_c^C$ is to  shift the $I_J$ response along the $V_C$ axis and
to slightly reduce the amplitude of the $\pi$-transition.}
\label{fig:delta}
\end{figure}
\section{PERFORMANCE I: ROLE OF MATERIALS CHOICE}
It is interesting to analyze the role of different superconductor
combinations, i.e., different ratios $T_c ^C /T_c ^J$. This effect
is shown in Fig. \ref{fig:delta} where the supercurrent vs bias
voltage (normalized to $\Delta_J /e$) characteristic is displayed
for different combinations of superconductors at $T=0.02\,T_c^J$.
\begin{figure}[ht!]
\begin{center}
\includegraphics[width=8.0cm,clip]{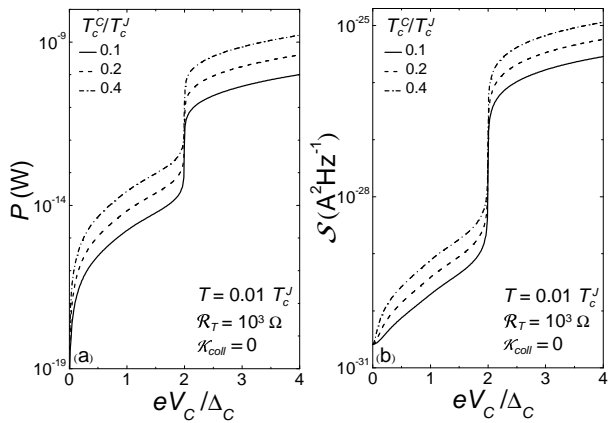}
\end{center}
\caption{(a) Power dissipated in the SINIS line vs $V_C$
calculated for various ratios $T_c ^C /T_c ^J$ and $T =0.01 T_c
^J$. (b) Noise power $\mathcal{S}$ vs $V_C$ calculated for the
same parameters as in (a). In all these calculations we set $T_c
^J =9.26$ K corresponding to Nb and $\mathcal{R}_T =1\,k\Omega$.}
\label{fig:power}
\end{figure}
The main effect is a shift in the $I_J$ response along the $V_C$
axis towards lower bias voltages upon reducing the critical
temperature $T_c^C$. Furthermore, the shape of the characteristics
appears to be virtually independent of the $T_c^C$ value, although
 a small decrease of the $\pi$-transition amplitude  gradually  develops
by decreasing the ratio $T_c ^C /T_c ^J$. Such a ratio can be
chosen to move the range of relevant $V_C$ towards lower values
and thus decrease the power dissipation $P=I_C V_C$, where $I_C$
is the control current across the SINIS channel. The function
$P(V_C)$ is plotted in Fig.~\ref{fig:power}(a) for some ratios
$T_c ^C /T_c ^J$ at $T=0.01 \,T_c ^J$, assuming $\mathcal{R}_T
=10^3 \,\Omega$ and $T_c ^J =9.26$ K (corresponding to Nb). The
impact of $\Delta_C$ in controlling power dissipation is easily
recognized. These effects clearly point out that $\Delta _C \ll
\Delta _J$ is the condition to be fulfilled in order to minimize
$P$. In practice, the power dissipation for $eV_C > 2\Delta$
constitutes an experimental problem as this energy needs to be
carried out from the reservoirs.

In a similar way the noise properties of the system are sensitive
to the different $T_c ^C /T_c ^J$ ratios. Assuming that the noise
through one junction is essentially uncorrelated from the noise
through the other, it follows that the input noise power
$\mathcal{S}$ in the control line can be expressed as
\begin{equation}
\mathcal{S}(V_C)=\frac{1}{\mathcal{R}_T} \int
_{-\infty}^{\infty}dE\mathcal{N}_1
[f(E)(1-\mathcal{F}_1)+\mathcal{F}_1(1-f(E))].
\end{equation}
The function $\mathcal{S}(V_C)$ is shown in
Fig.~\ref{fig:power}(b) for the same parameters as in
Fig.~\ref{fig:power}(a). For example, for $T_c ^C /T_c ^J =0.1$
(corresponding roughly to the combination Al/Nb), $P$ obtains
values of the order of a few  fW and $\mathcal{S}$ of some
$10^{-30}$ A$^2$Hz$^{-1}$ in the cooling regime, while these
values are enhanced respectively to  few tens of pW and $10^{-26}$
A$^2$Hz$^{-1}$ for biases around the $\pi$-transition.

\begin{figure}[ht!]
\begin{center}
\includegraphics[width=8.0cm,clip]{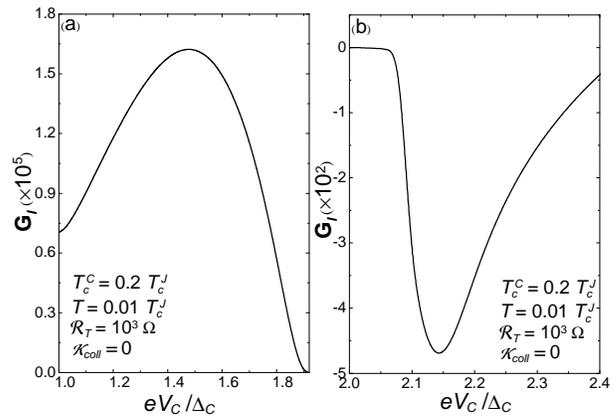}
\end{center}
\caption{Differential current gain $\mathbf{G}_I$ vs $V_C$ for
$T_c ^C /T_c ^J =0.2$ and $T=0.01\,T_c^J$: in (a) $\mathbf{G}_I$
is shown for $V_C < V_C^\ast$ while in (b) it is shown in the
high-bias region. In these calculations we set
$\mathcal{K}_{coll}=0$ and $T_c ^J =9.26$ K (Nb).}
\label{fig:Igain}
\end{figure}
\section{PERFORMANCE II: GAIN AND SWITCHING TIME}
In light of the possible use of this operational principle for
device implementation, let us comment on the available gain and
switching times. Input ($V_{in}=V_C$)  and output ($V_{out}=I_J
R_N$) voltages are of the order of $\Delta _C /e$ and $E_{Th}^J
/e$, respectively, so that the voltage gain is
$\mathbf{G}_{V}=V_{out}/V_{in} \approx E_{Th}^J/\Delta_C$. As
$\Delta_C \rightarrow E_{Th}^J$, the device characteristics become
smoother (see Sec.~\ref{sec:ldep}) and the control becomes less
efficient, so that it is hardly possible to achieve voltage gain.
On the other hand, differential current gain $\mathbf{G}_I =dI_J
/dI_C$ can be very large. For $V_C
>V_C ^\ast$, a simple estimate  gives
\begin{equation}
\mathbf{G}_I \sim (E_{Th}^J/\Delta _C)(\mathcal{R}_T /R_N),
\end{equation}
meaning that with realistic ratios $\mathcal{R}_T /R_N$ ($\sim
10^3$), $\mathbf{G}_I$ can
 exceed $10^2$.
$\mathbf{G}_I (V_C)$ can be calculated noting that $\mathbf{G}_I
=(\frac{dI_J}{dV_C})(\frac{dI_C}{dV_C})^{-1}$. In
Fig.~\ref{fig:Igain} we plot $\mathbf{G}_I$ for $T_c ^C /T_c ^J
=0.2$. This calculation reveals that $\mathbf{G}_I$ can reach huge
values, with some $10^{5}$ for $V_C <V_C ^\ast$ \cite{gain} and
several $10^{2}$ in the opposite regime. Remarkably, gain is
almost unchanged also in the presence of weak inelastic scattering
(i.e., $\mathcal{K}_{coll}=1$). The same holds for $P$ and
$\mathcal{S}$.

Operation of the Josephson junction in the \textit{dissipative}
regime makes it possible to exploit the system for power gain. Let
us assume that the voltage across the Josephson junction ($V_J$)
is fixed and that the current-voltage characteristic is
nonhysteretic. According to the  RSJ (resistively shunted
junction) model \cite{RSJ} for overdamped junctions, the voltage
drop across the junction at $T=0$ reads
\begin{equation}
V_J=R_N \sqrt{I^2-I_{Jc}^2}, \label{rsj}
\end{equation}
where $I_{Jc}$ is the critical current and $I$ is the total
current through the Josephson junction. The differential power
gain is defined as $\mathbf{G}_P =dP_{out}/dP_{in}$ where
$P_{out}=V_J I $, $P_{in}=V_C I_C$. Using Eq. (\ref{rsj}) one
obtains
\begin{align}
\mathbf{G}_P &= V_J \frac{d I}{d(V_C I_C)} = \frac{V_J
I_{Jc}}{\sqrt{\left(\frac{V_J}{R_N}\right)^2+I_{Jc}^2}}
\frac{dI_{Jc}}{dI_C} \frac{dI_C}{d(V_C I_C)} \\&=
\frac{I_{Jc}}{\sqrt{\left(\frac{V_J}{R_N}\right)^2+I_{Jc}^2}}
\frac{V_J}{V_C+I_C R_{\rm SINIS}(V_C)} \mathbf{G}_I,
\end{align}
where $R_{\rm SINIS} = dV_C/dI_C$ is the differential resistance
of the SINIS line at the operating current $I_C$. Choosing
$V_J=I_{JC} R_N$ we obtain
\begin{equation}
\mathbf{G}_P = \frac{V_J}{\sqrt{2}(V_C+I_C R_{\rm SINIS})}
\mathbf{G}_I \ge \frac{V_J}{2\sqrt{2} V_C} \mathbf{G}_I.
\label{po}
\end{equation}
The thermal smearing of Eq.~(\ref{rsj}) \cite{RSJ} will somewhat
decrease $\mathbf{G}_P$, but not by orders of magnitude. A further
simplification of Eq. (\ref{po}) can be additionally given,
recalling that $V_J \sim E_{Th}^J/e$ and $V_C \sim 2\Delta
_{C}/e$, so that
\begin{equation}
\mathbf{G}_P \sim \frac{E_{Th}^{J}}{4\sqrt{2}\Delta_C}\mathbf{G}_I
\simeq \mathbf{G}_V \mathbf{G}_I. \label{gainsimpl}
\end{equation}
A straightforward estimate for the differential power gain from
Eq. (\ref{gainsimpl}) gives $\mathbf{G}_P \sim 10^2\div 10^3$ for
$V_C <V_C^\ast$ and $\sim 10$ for $V_C
>V_C^\ast$. This means that, in the suitable operating mode, the
present structure is able to provide large power gain.

As a final issue we want to briefly address the time scales
intrinsic to this mesoscopic device when it is operated in the
dynamic regime. The highest operating frequency $\nu$ of the
transistor is limited by the smallest energy in the system: $\nu
\leq \min \frac{1}{h}\{\Delta_C ,\Delta_J ,E_{Th}^C ,E_{Th}^J ,
h(\mathcal{R}_T \mathcal{C})^{-1},h e/I_C\}$ where $\mathcal{C}$
is the tunnel junction capacitance. For an optimized device,
working frequencies of the order of $10^{11}$ Hz can be
experimentally achieved in the high-voltage regime $V_C > V_C
^\ast$. For $V_C <V_C ^\ast$, conversely, the response is slower
(of the order of some $10^8$ Hz), owing to the large value of
subgap resistance that becomes the speed-limiting factor due to
the long discharging time through the junctions. However, this
analysis relies on the fact that the relaxation time scales in the
superconducting reservoirs in the control part of the structure
are much shorter than the above scales. These scales depend on the
geometry of the system (the real size of the reservoirs), but in
practice, they may seriously limit the speed of the device unless
special care is taken. We stress that the detailed prediction of
the transistor dynamic response would require a separate analysis
that is beyond the scope of the present paper.

\section{CONCLUSIONS}
In this paper we have shown that superconductivity-induced
nonequilibrium can be used to finely control the supercurrent
flowing through a SNS Josephson junction. The peculiar
quasiparticle energy distribution, realized in the normal island
of a SINIS control line when a bias voltage is applied between the
superconducting electrodes, allows us to drive the SNS junction
into different transistor regimes. With increasing control
voltage, we have found that the supercurrent first shows a steep
enhancement, then undergoes a $\pi$-transition and finally decays
to zero for larger voltages. Furthermore, we have analyzed in
detail the effect of inelastic scattering within the normal
island, the dependence of the supercurrent on the Josephson
junction length, the effect of the presence of lateral arms on the
spectral supercurrent, and the behavior of the supercurrent
harmonics. Finally we have addressed the optimization of the
device, both in terms of material choice and characteristic
figures, such as current gain, power gain and operating
frequencies. In view of these last points, we wish to point out
that such a device could be successfully exploited as a first
amplification stage in SQUID-based cryogenic electronics.

\section*{ACKNOWLEDGMENTS}
We thank A. Anthore, M. H. Devoret, K. K. Likharev, F. Pierre, L.
Roschier, A. M. Savin, and V. Semenov for helpful discussions.
This work was supported in part by MIUR under the FIRB project
RBNE01FSWY and by the EU (RTN-Nanoscale Dynamics).


\begin{thebibliography}{9}

\bibitem{articles}
    See, for example, \emph{Theory of Nonequilibrium Superconductivity},
    N.B. Kopnin (Clarendon Press, Oxford, 2001).
\bibitem{baselmans}
    J.J.A. Baselmans, A.F. Morpurgo, B.J. van Wees, and T.M. Klapwijk, Nature (London) {\bf 397}, 43
    (1999).

\bibitem{baselmansprb}
    J.J.A. Baselmans, B.J. van Wees, and T.M. Klapwijk, Phys. Rev. B {\bf 63}, 094504
    (2001).

\bibitem{huang}
    J. Huang, F. Pierre, T.T. Heikkil\"{a}, F.K. Wilhelm, and N.O. Birge, Phys. Rev. B {\bf 66}, 020507
    (2002).

\bibitem{tak}
    R. Shaikhaidarov, A.F. Volkov, H. Takayanagi, V.T. Petrashov, and P. Delsing, Phys. Rev. B {\bf
    62}, R14649 (2000).

\bibitem{baselmansphase}
    J.J.A. Baselmans, T.T. Heikkil\"{a}, B.J. van Wees, and T.M. Klapwijk, Phys. Rev. Lett. {\bf 89}, 207002 (2002).

\bibitem{volkov}
    A. F. Volkov, Phys. Rev. Lett. {\bf 74}, 4730 (1995).

\bibitem{wilhelm}
    F.K. Wilhelm, G. Sch\"on, and A.D. Zaikin, Phys. Rev. Lett. {\bf 81},
1682 (1998).

\bibitem{yip}
    S.-K. Yip, Phys. Rev. B {\bf 58}, 5803 (1998).

\bibitem{pothier}
    H. Pothier, S. Gu\'{e}ron, N.O. Birge, D. Esteve, and M.H. Devoret,  Phys. Rev. Lett. {\bf 79}, 3490 (1997).

\bibitem{SINISTrans}
    F. Giazotto, T.T. Heikkil\"{a}, F. Taddei, R. Fazio, J.P.
    Pekola, and F. Beltram, Phys. Rev. Lett. {\bf 92},
    137001 (2004).

\bibitem{belzig}
    See W. Belzig, F.K. Wilhelm, C. Bruder, G. Sch\"on, and A.D. Zaikin,  Superlattices Microstruct. {\bf 25},
    1251 (1999) and references therein.

\bibitem{tero}
    T.T. Heikkil\"{a}, J. S\"{a}rkk\"{a}, and F.K. Wilhelm,  Phys. Rev. B {\bf 66}, 184513 (2002).


\bibitem{usadel}
    K.D. Usadel,  Phys. Rev. Lett. {\bf 25}, 507 (1970).


\bibitem{Heslinga}
    D.R. Heslinga, and T.M. Klapwijk, Phys. Rev. B {\bf 47},
5157 (1993).

\bibitem{brinkman03}
    A. Brinkman, A.A. Golubov, H. Rogalla, F.K. Wilhelm, and M.Yu. Kupriyanov,  Phys. Rev. B {\bf 68}, 224 513 (2003).

\bibitem{lowenote}
    At low control voltages there is a region of energies where
$\mathcal{N}_1=\mathcal{N}_2=0$. We assume that there the
(otherwise weak) coupling to phonons makes the distributions at
those energies equal to the equilibrium Fermi distribution. A more
detailed discussion and another type of coupling is given in
Ref.~\onlinecite{cooler}.

\bibitem{baselmansthesis}
    A similar behavior was predicted by J.J.A. Baselmans, Ph. D. thesis, University of Groningen (2002).

\bibitem{cooler}
    J.P. Pekola, T.T. Heikkil\"{a},A.M. Savin, J.T. Flyktman, F. Giazotto, and F.W.J.
    Hekking,  Phys. Rev. Lett. {\bf 92}, 056804 (2004).

\bibitem{leivo}
    M.M. Leivo, J.P. Pekola, and D.V. Averin,  Appl. Phys. Lett. {\bf 68},
1996 (1996).

\bibitem{intspec} The interaction effects are not taken into
account in the spectral supercurrent --- this approximation worked
fine in Refs.~\onlinecite{baselmansprb,huang}.

\bibitem{anthore}
    A. Anthore, F. Pierre, H. Pothier, and D. Esteve,  Phys. Rev. Lett. {\bf 90}, 076806 (2003).

\bibitem{nag}
    K.E. Nagaev, Phys. Rev. B {\bf 52}, 4740 (1995).


\bibitem{AA}
    B.L. Altshuler and A.G. Aronov,  Zh. Eksp. Teor. Fiz. {\bf 75},
    1610 (1978) [ Sov. Phys. JETP {\bf 48}, 812 (1978)].

\bibitem{kamenev}
    A. Kamenev and A. Andreev,  Phys. Rev. B {\bf 60}, 2218 (1999).

\bibitem{huard}
    B. Huard, A. Anthore, F. Pierre, H. Pothier, N.O. Birge, D.
    Esteve, preprint  cond-mat/0404208.

\bibitem{giazotto}
    F. Giazotto, F. Taddei, T.T. Heikkil\"{a}, R. Fazio, and F. Beltram,   Appl. Phys. Lett. {\bf 83}, 2877 (2003).

\bibitem{device}
     This is straightforward assuming as typical parameters  $\mathcal{R}_T=10^3$ $\Omega$,
    $D=0.02$ m$^2$/s and $\Delta_C =200$ $\mu$eV.
\bibitem{gain}
    In this calculation we chose to include depairing by a phenomenological, but realistic parameter $\Gamma = 10^{-4} \Delta_C$ \cite{cooler}.
Its omission would lead to extremely high values of
$\mathbf{G}_I$. For $V_C \ll V_C ^\ast$ and low temperature,
$\Gamma$ determines the differential resistance $R_{SINIS}\approx
\mathcal{R}_T \Delta_C /\Gamma$ of the SINIS line. Qualitatively,
the large $\mathbf{G}_I$ stems partially from the fact $R_{SINIS}
\gg R_N$.

\bibitem{RSJ}
    V. Ambegaokar and B.I. Halperin, Phys. Rev. Lett. {\bf
    22}, 1364 (1969).

\end{thebibliography}
\end{document}